\documentclass[english,11pt]{extarticle}
\usepackage[T1]{fontenc}
\usepackage[latin9]{inputenc}
\usepackage{babel}
\usepackage{geometry}
\usepackage{amsthm}
\usepackage{amsmath}

\usepackage{relsize}
\usepackage{amssymb}
 \usepackage{amsfonts}
 \usepackage{amscd}
 \usepackage[all]{xy}
\usepackage[dvipsnames]{xcolor}
\usepackage{pdfsync}
\usepackage{url}
\usepackage[sc]{mathpazo}

\usepackage[ruled, vlined]{algorithm2e}
\usepackage{multirow}
\usepackage{color}
\newcommand{\LX}{{{\mathcal L}_X}}
\newcommand{\DD}{{{\mathcal D}_X}}
\newtheorem{remark}{\bf Remark}
\newtheorem{notation}{\bf Notation}
\newtheorem{lemma}{\bf Lemma}
\newtheorem{theorem}{\bf Theorem}

\newtheorem{corollary}{\bf Corollary}
\newtheorem{proposition}{\bf Proposition}
 \newtheorem{defn}{\bf Definition}

 \newfont{\mycrnotice}{ptmr8t at 7pt}
\newfont{\myconfname}{ptmri8t at 7pt}

\begin{document}

\title{Data-Discriminants of Likelihood Equations
\thanks{This research paper is partly supported by 2014 NIMS Thematic Program on Applied Algebraic Geometry.}
}

%%%%%%%%%%%%%%
\author{
Jose Israel Rodriguez
\thanks{
   {ACMS Department},
   {University of Notre Dame},
      {Notre Dame, IN 46556};
       {\tt jo.ro@nd.edu.}
       This material is based upon work supported by the National Science Foundation under Award No. DMS-1402545, as well as the hospitality of the Simons Institute.}
\and
Xiaoxian Tang
\thanks{
      {CAMP, National Institute for Mathematical Sciences},
       {Jeonmin-Dong, Yuseong-gu},
       {Daejeon, Republic of Korea 305-811};
     {\tt tangxiaoxian@nims.re.kr}}.% (Corresponding Author).
}

\maketitle
\begin{abstract}
Maximum likelihood estimation (MLE) is a fundamental computational problem in statistics.
The problem is to maximize the likelihood function with respect to given data on a statistical model.
An algebraic approach to this problem is to solve a very structured parameterized polynomial system called likelihood equations. 
For general choices of data, the number of complex solutions to the likelihood equations is finite and called the ML-degree of the model. 

The only solutions to the likelihood equations that are statistically meaningful are the real/positive solutions. However,   
the number of real/positive solutions is not characterized by the ML-degree. 
We use discriminants to classify data according to the number of real/positive solutions  of the likelihood equations.
We call these discriminants  data-discriminants (DD). 
We  develop a probabilistic algorithm for computing DDs. 
Experimental results show that, for the benchmarks we have tried,  
the probabilistic algorithm  
 is more efficient than the standard elimination algorithm. 
Based on the computational results, we  discuss the real root classification problem for the $3$ by $3$ symmetric matrix~model. 
\end{abstract}

%A category including the fourth, optional field follows...
%\category{G.3}{Probability and Statistics}{Statistical Computing}
%\category{I.1.2}{Computing Methodologies}{Algorithms}[Algebraic Algorithms]
%\category{D.2.8}{Software Engineering}{Metrics}[complexity measures, performance measures]

%\terms{Theory, Algorithms}

%\keywords{Maximum likelihood estimation, Likelihood equation, Data-Discriminant, Real root classification} % NOT required for Proceedings

\section{Introduction}\label{introduction}
We begin the introduction with an illustrative example.
 Suppose we have a weighted four-sided die such that 
the  {\em probability} $p_i$ of observing side $i$ $(i=0, 1, 2, 3)$ of the die satisfies the constraint $p_0+2p_1+3p_2-4p_3=0$. 
We toss the die 1000 times and record a $4$-dimensional {\em data} vector $(u_0, u_1, u_2, u_3)$, where $u_i$ is the number of times we observe the side $i$. 
 We want to determine the probability distribution $(p_0,p_1,p_2,p_3)\in\mathbb{R}_{>0}^4$ that best explains the data subject to the constraint. 
  One approach is by {\em maximum likelihood estimation} (MLE):
\begin{center}
{\bf Maximize the {\em likelihood function}
${p_{0}^{u_{0}}p_{1}^{u_{1}}p_{2}^{u_{2}}p_{3}^{u_{3}}} $ subjected to}
\footnotesize{\[p_0+2p_1+3p_2-4p_3=0,
p_{0}+ p_{1}+p_{2}+ p_{3}=1, \]
\[ p_{0}>0, p_{1}>0, p_{2}>0,  \text{and}\;  p_{3}>0.\]}
\end{center}
%where 
%\begin{equation*}
%p_{ij}=
%\begin{cases}
%\overline{p}_{ij} & i=j\\
%\frac{1}{2}\overline{p}_{ij} & i<j
%\end{cases}, \;\;
%u_{ij}=
%\begin{cases}
%\overline{u}_{ij} & i=j\\
%\overline{u}_{ij}+\overline{u}_{ji} & i<j
%\end{cases}.
%\end{equation*}

For some statistical models,  the  MLE  problem can be solved by  well known {\em hill climbing} algorithms such as the EM-algorithm. However, the hill climbing method can fail if there is more than one local maximum.  
 Fortunately, it is known  that the MLE problem can be solved by solving the system of  {\em likelihood equations} \cite{SAB2005, CHKS2006}:
  \begin{align*}
F_0&=p_{0}\lambda_1+p_{0}\lambda_2-  u_{0}& F_3&=p_{3}\lambda_1-4p_{3}\lambda_2-   u_{3}\\
F_1&=p_{1}\lambda_1+2p_{1}\lambda_2 -  u_{1}& F_4&=p_0+2p_1+3p_2-4p_3\\
F_2&=p_{2}\lambda_1+3p_{2}\lambda_2-  u_{2}& F_5&=p_{0} + p_{1} +p_{2}+p_{3} - 1
\end{align*}
%  \begin{align*}
%F_0&=p_{0}\lambda_1+p_{0}\lambda_2-  u_{0}=0\\
%F_1&=p_{1}\lambda_1+2p_{1}\lambda_2 -  u_{1}=0\\
%F_2&=p_{2}\lambda_1+3p_{2}\lambda_2-  u_{2}=0\\
%F_3&=p_{3}\lambda_1-4p_{3}\lambda_2-   u_{3}=0\\
%F_4&=p_0+2p_1+3p_2-4p_3=0\\
%F_5&=p_{0} + p_{1} +p_{2}+p_{3} - 1=0
%\end{align*}
where $\lambda_1$ and $\lambda_2$ are newly introduced indeterminates (Lagrange multipliers) for formulating the likelihood equations. More specifically, for given $(u_{0},  u_{1},  u_{2},  u_{3})$,  if $(p_{0},p_{1},p_{2},p_{3})$ is a critical point of the likelihood function, then there exist complex numbers $\lambda_1$ and $\lambda_2$ such that $(p_{0}, p_{1},p_{2}, p_{3}, \lambda_1, \lambda_2)$ is a solution of the polynomial system.  
For randomly chosen data $u_{i}$, the  likelihood equations have $3$ complex solutions. However, only  solutions with positive coordinates $p_{i}$
are statistically meaningful. A solution with all positive $p_i$ coordinates is said to be a positive solution.   So an important problem is  {\em real root classification} (RRC):
\begin{center}
{\bf For which $u_{i}$, the polynomial system has $1, 2$ and $3$
real/positive solutions? }
\end{center}

According to the theory of computational (real) algebraic geometry \cite{BP2001, DV2005},  the number of (real/positive) solutions only changes when the
data $u_{i}$ goes across some ``special'' values (see Theorem \ref{theorem2}). The set of  ``special" $u_{i}$ is a {\em (projective)} {\em variety} (see Lemma 4 in \cite{DV2005}) in  ($3$ dimensional complex projective space)
$4$-dimensional complex space.   
%{\color{red} Is this confusing because we switch from complex space to real space. }
The number of real/positive solutions is uniform over each open connected component determined by the variety. In other words, the ``special'' $u_{i}$  plays the similar role as the {\em discriminant} for univariate  polynomials. 
 The first step of  RRC is calculating the ``special'' $u_i$, leading to the discriminant problem:
\begin{center}
{\bf How to effectively compute the ``special'' $u_{i}$?} 
\end{center}

Geometrically, the ``special'' $u_{i}$ is a projection of a variety. So in principle, it can be computed by {\em elimination} ( see Chapter 3, page 115--128 in \cite{CLO2007}).
For instance, by the command {\tt eliminate} in {\tt Macaulay2} \cite{M2}, we  compute that the ``special" $u_{i}$ in the illustrative example form a hypersurface defined by  a homogenous polynomial in $(u_0, u_1, u_2, u_3)$ (see  Example 1).
However, for most MLE problems,  due to the large size of likelihood equations, the elimination computation is too expensive.  
In this paper, we discuss the ``discriminant" problem for the likelihood equations.  The contributions of the paper are listed as follows. 

\textbullet ~For likelihood equations,  we show that the ``special" $u_i$ form a projective variety.  We call the homogenous polynomial that generates the codimension $1$ component of the projective variety the  {\em data-discriminant}. This name  distinguishes it from the  {\em weight-discriminant} for the likelihood equations (which replaces the condition $p_0+\cdots+p_n=1$ with the condition  
$h_0p_0+\cdots+h_np_n=1$ with 
parameters $h_0,\dots,h_n$). 
%We show that different formulations of the likelihood equations yield different data discriminants. 

\textbullet  ~For algebraic statistical models, we develop a probabilistic algorithm to compute data-discriminants. We implement the algorithm in {\tt Macaulay2}. Experimental results show that the probabilistic algorithm  is more efficient than the standard elimination algorithm. 

\textbullet  ~We discuss the real root classification for the $3\times 3$ symmetric matrix model, which inspire  future work. 

We remark that our work can be viewed as following the numerous efforts in applying computational algebraic geometry to tackle MLE and critical points problems \cite{SAB2005, CHKS2006, BHR2007, HS2010,Uhler2012, GDP2012, FES2012, EJ2014, HS2014, HRS,Rod14}.  

The paper is organized as follows. 
The formal definition of the data-discriminant is introduced in  Section \ref{definition}. The standard elimination algorithm and the probabilistic algorithm  are presented in  Section \ref{algorithm}. Experimental results  comparing the two algorithms are shown in  Section \ref{experiment}.  The  real root classification  of the $3\times 3$ symmetric matrix model  and  conclusion are   given in Section \ref{conclusion}. 
\section{Definition}\label{definition}
In this section, we discuss how to define ``data-discriminant''.  We assume the readers are familiar with  elimination theory  (see Chapter 3 in \cite{CLO2007}).
 
\begin{notation}
%{\bf Notation 1.} 
%${\mathbb Q}$, ${\mathbb R}$, ${\mathbb C}$ denote the {\em fields of rational, real and complex numbers}, respectively. 
Let ${\mathbb P}$ denote the  {\em projective closure} of the complex numbers ${\mathbb C}$.
%\end{notation}
%\begin{notation}
%{\bf Notation 2.}
For  homogeneous polynomials $g_1,\ldots,g_s$ in  ${\mathbb Q}[p_0, \ldots,p_n]$, 
${\mathcal V}(g_1,\ldots,g_s)$ denotes the {\em projective variety} in ${\mathbb P}^n$ defined by $g_1, \ldots, g_s$.
%\end{notation}
%\begin{notation}
%{\bf Notation 3.}
Let $\Delta_{n}$ denote the $n$-dimensional {\em probability simplex} 
$\{(p_0, \ldots, p_n)\in {\mathbb R}^{n+1}|p_0>0, \ldots,p_n>0, p_0+\cdots+p_n=1\}$. 
\end{notation}

\begin{defn}\cite{SAB2005}{\bf (Algebraic Statistical Model and Model Invariant)}
%{\bf Definition 1. (Algebraic Statistical Model and Model Invariant)}
The set $X$ is said to be an {\em algebraic statistical
model} if $X={\mathcal V}(g_1,\ldots,g_s)\cap \Delta_{n}$  where $g_1, \ldots, g_s$ 
define an irreducible generically reduced projective variety. 
Each $g_k$ $(1\leq k\leq s)$ is said to be a {\em model invariant} of~$X$. 
\end{defn}

For a given algebraic statistical model, there are several different ways to formulate the likelihood equations \cite{SAB2005}. In this section, we  introduce the Lagrange likelihood equations and  define the data-discriminant for this formulation. One can similarly define data-discriminants for other formulations of the likelihood equations. 

\begin{notation}
%{\bf Notation 4.}
For any $f_1, \ldots, f_m$ in the polynomial ring ${\mathbb Q}[x_1, ..., x_k]$, ${\mathcal V}_a(f_1, \ldots, f_m)$ denotes the {\em affine variety} in ${\mathbb C}^k$ defined by $f_1, \ldots, f_m$ and 
$\langle f_1, \ldots, f_m \rangle$ denotes the  {\em ideal} generated by $f_1, \ldots, f_m$.
For an ideal  $I$  in ${\mathbb Q}[x_1, \ldots, x_k]$,  ${\mathcal V}_a(I)$ denotes the {\em affine variety} defined by $I$. 

\end{notation}

\begin{defn}\label{LE}\cite{EJ2014}{\bf (Lagrange Likelihood Equations and Correspondence)}
%{\bf Definition 2. (Lagrange Likelihood Equations and Correspondence)}
Given an algebraic statistical model $X$. The system of polynomial equations below is said to be  the  {\em Lagrange likelihood
equations} of $X$:  
%\begin{equation}\label{LLE}
\begin{align*}
&F_{0}=\;\;\;\;p_0(\lambda_1+\frac{\partial g_1}{\partial p_0}\lambda_2+\cdots+\frac{\partial g_s}{\partial p_0}\lambda_{s+1})-u_0=0\\
%\end{align*}
%\begin{align*}
&\cdots\\
%\end{align*}
%\begin{align*}
&F_{n}=\;\;\;\;p_n(\lambda_1+\frac{\partial g_1}{\partial p_n}\lambda_2+\cdots+\frac{\partial g_s}{\partial
p_n}\lambda_{s+1})-u_n=0
\end{align*}
\begin{align*}
&F_{n+1}=\;\;\; \;g_1(p_0, \ldots,p_n)=0\\
%\end{align*}
%\begin{align*}
&\cdots\\
%\end{align*}
%\begin{align*}
&F_{n+s} =\;\; \;\;g_s(p_0,\ldots,p_n)=0\\
%\end{align*}
%\begin{align*}
&F_{n+s+1}=\;\;\;\;p_0+\cdots+p_n-1=0
\end{align*}
%\end{equation}
where 
%\begin{itemize}
%\item 
$g_1,\ldots,g_s$ are the model invariants of $X$ and
$u_0, \ldots, u_n$, $p_0, \ldots,p_n$, $\lambda_1,\ldots,\lambda_{s+1}$ are indeterminates
(also denoted by ${\mathbf u}$, ${\mathbf p}$, ${\Lambda}$).
More specifically, 
     %\begin{itemize} 
        
             %\item 
            ~~\textendash~$p_0,\ldots,p_n, \lambda_1,\ldots,\lambda_{s+1}$ are unknowns,
             %\item 
             
             ~~\textendash~$u_0,\ldots,u_n$~are parameters.
     % \end{itemize}
%\end{itemize}

\noindent
${\mathcal V}_a(F_0, \ldots, F_{n+s+1})$, namely the set  %{\color{red} Do we need a closure for the set below??}
\[\{({\mathbf u}, {\mathbf p}, {\Lambda})\in {\mathbb C}^{n+1}\times {\mathbb C}^{n+1}\times {\mathbb C}^{s+1}|F_0=0, \ldots, F_{n+s+1}=0\},\]
%$({\bf u}, {\bf p}, {\Lambda})$ in ${\mathbb C}^{2n+s+3}$ such that $F_1=0, \ldots, F_{n+s+1}=0$,
 is said to be the {\em Lagrange likelihood correspondence} of $X$ and 
denoted by $\LX$. 
\end{defn}

\begin{notation}
%{\bf Notation 5.}
Let $\pi$ denote the  {\em canonical projection} from the ambient space of the Lagrange likelihood correspondence  to the $\mathbb{C}^{n+1}$ associated to the ${\mathbf u}$ indeterminants $\pi$: 
 ${\mathbb C}^{n+1}\times {\mathbb
C}^{n+s+2}\rightarrow{\mathbb C}^{n+1}$.
\end{notation}

Given an algebraic statistical model $X$ and a data vector ${\bf u}\in {\mathbb R}_{>0}^n$,  the {\em maximum likelihood estimation} (MLE) problem is to
%\noindent  
%{\bf MLE Problem.}
%\begin{center}
{\bf maximize the} {\em likelihood function}
{\bf $p_0^{u_0}\cdots p_n^{u_n}$ subject to $X$.}
%\end{center}
The MLE problem can be solved by computing $\pi^{-1}({\mathbf u})\cap \LX$.
 More specifically, if ${\mathbf p}$ is a regular point of ${\mathcal V}(g_1,\ldots,g_s)$, then ${\mathbf p}$ is a critical point of the likelihood function if and only if there exist $\Lambda\in {\mathbb C}^{s+1}$ such that $({\mathbf u}, {\mathbf p}, {\Lambda})\in \LX$.  Theorem 1
 %\ref{MLD} 
 states that for  a general data vector ${\mathbf u}$,  $\pi^{-1}({\mathbf u})\cap \LX$
 is a finite set and the cardinality of $\pi^{-1}({\mathbf u})\cap \LX$ is constant over a dense Zariski open set, which inspires the definition of ML-degree.  For details, see \cite{SAB2005}.

%{\bf Notation 6.}
%\begin{notation}
%For any finite set $S$, $\#  S$ denotes the {\em cardinality} of $S$.
%\end{notation}

\begin{theorem}\label{MLD}\cite{SAB2005}
%{\bf Theorem 1.}
For an algebraic statistical model $X$, 
 there exist an affine variety $V\subset {\mathbb C}^{n+1}$ and a non-negative integer $N$  such that for any ${\mathbf u}\in {\mathbb
C}^{n+1}\backslash V$,  
\[\#\pi^{-1}({\mathbf u})\cap \LX = N.\] 
 \end{theorem}

\begin{defn}\cite{SAB2005}{\bf (ML-Degree)}
%[ML-Degree]
%{\bf Definition 3. (ML-Degree)}
For an algebraic statistical model $X$, the non-negative integer $N$ stated in Theorem \ref{MLD}
is said to be the {\em ML-degree} of $X$. 
\end{defn}

\begin{notation}
%{\bf Notation 7.}
For any $S$ in  ${\mathbb C}^{n+1}$, ${\mathcal I}(S)$ denotes the ideal 
\[\{D\in {\mathbb Q}[{\mathbf u}]|D(a_0, \ldots, a_n)=0, \forall (a_0, \ldots, a_n)\in S\}.\]
$\overline{S}$ denotes the {\em affine closure} of $S$ in ${\mathbb C}^{n+1}$, namely ${\mathcal V}_a({\mathcal I}(S))$.
\end{notation}

\begin{defn}\label{nddv}
For an algebraic statistical model $X$, suppose $F_0, \ldots, F_{n+s+1}$ are defined as  in Definition \ref{LE}. 
Let 
%\begin{itemize}
%\item 
%\textbullet~$\LX^{{\mathbb P}}$ denotes the $({\bf p}, {\Lambda})$-{\em projective closure} of $\LX$ in the space  ${\mathbb C}^{n+1}\times {\mathbb P}^{n+s+2}$, namely the variety defined by the 
%$({\bf p}, \Lambda)$-{\em homogenization} of  $\langle F_0, \ldots, F_{n+s+1}\rangle$;
%\item 
%${\mathbb H}_{\infty}$ denotes the space $({\mathbb C}^{n+1}\times {\mathbb P}^{n+s+2})\backslash {\mathbb C}^{2n+s+3}$. 
 %\textbullet~
 $J$ denote {\footnotesize\[\det \left[
\begin{matrix}
\frac{\partial F_0}{\partial p_0} & \cdots & \frac{\partial
F_0}{\partial
p_n} & \frac{\partial F_{0}}{\partial \lambda_1} & \cdots & \frac{\partial F_{0}}{\partial \lambda_{s+1}}\\
\vdots & \ddots & \vdots &\vdots & \ddots & \vdots \\
\frac{\partial F_{n+s+1}}{\partial p_0} & \cdots & \frac{\partial
F_{n+s+1}}{\partial
p_n}& \frac{\partial F_{n+s+1}}{\partial \lambda_1} & \cdots & \frac{\partial F_{n+s+1}}{\partial \lambda_{s+1}}
\end{matrix}
\right]%_{(n+s+1)\times (n+s+1)}
.\]}
%\end{itemize}

\noindent
Then, we have the following:

%\begin{itemize}
%\item 
\textbullet~$\LX_{\infty}$ denotes the {\em set of non-properness} of $\pi$, i.e., the set of the $u\in \overline{\pi(\LX)}$ such that there does not exist 
a   compact neighborhood $U$ of $u$ where
$\pi^{-1}(U)\cap \LX$ is  compact;
%\textbullet~$\LX_{\infty}$ denotes
 %$\pi(\LX^{{\mathbb P}}\cap ({\mathbb C}^{n+1}\times {\mathbb P}^{n+s+2})\backslash {\mathbb C}%^{2n+s+3})$;
%$\pi(\LX^{{\mathbb P}}\cap {\mathbb H}_{\infty})$, 
%\item
 
\textbullet~$\LX_{J}$ denotes
$\overline{\pi(\LX\cap {\mathcal V}_a(J))}$;
%\item 

\textbullet~$\LX_{p}$ denotes $\overline{\pi(\LX\cap {\mathcal V}_a(\Pi_{k=0}^np_k))}$.
%\end{itemize}
\end{defn}
%\begin{remark}

The geometric meaning of $\LX_{p}$ and $\LX_{J}$ are as follows. 
The first, $\LX_{p}$, is the projection of the intersection of the Lagrange likelihood correspondence with the  coordinate hyperplanes. 
The second, $\LX_{J}$, is the projection of the intersection of the Lagrange likelihood correspondence with the hypersurface defined by $J$. Geometrically, $\LX_{J}$ is the closure of the union of the projection of the singular
locus of $\LX$ and the set of
critical values of the restriction of $\pi$ to the regular locus of $
\LX$ (see Definition 2 in \cite{DV2005}).
%{\color{magenta} Is this Bernd's definition of discriminant for likelihood equations in Likelihood geometry?}. 

The Lagrange likelihood equations define an affine variety.  As we continuously deform the parameters $u_i$,  coordinates of a solution can tend to infinity. 
Geometrically,   $\LX_{\infty}$ is the set of the data ${\mathbf u}$ such that the Lagrange likelihood equations have some solution $({\mathbf p}, \Lambda)$ at infinity; 
this is the closure of the set of ``non-properness''  as defined in the page 1, \cite{Jelonek1999} and page 3, \cite{DS2004}.
  It is known that the set of non-properness of $\pi$ is closed and can be computed by Gr\"obner bases (see Lemma 2 and Theorem 2 in \cite{DV2005}).  
%\end{remark}
%\begin{remark}
%{\bf Remark 1.}
%It is well known \cite{M1976, CLO2007, DV2005} that $\pi(\LX^{{\mathbb P}}\cap {\mathbb H}_{\infty})=\overline{\pi(\LX^{{\mathbb P}}\cap {\mathbb H}_{\infty})}$. 
%\end{remark}

The ML-degree encaptures geometry of the likelihood equations  over the complex numbers. 
However, statistically meaningful solutions  occur over real numbers. 
Below, Theorem \ref{theorem2}  states that  $\LX_{\infty},$  $\LX_{J}$ and 
$\LX_{p}$ define open connected components such that the number of real/positive solutions is uniform over each open connected component. 
 Theorem 2 is a corollary of \textit{Ehresmann's theorem} for which there exists semi-algebraic statements since 1992 \cite{CS1992}.

\begin{theorem}\label{theorem2}
%\cite{M1976, BPR2006}
%{\bf Theorem 2.}
For any algebraic statistical model~$X$, 
%\begin{itemize}

%\item 
\textbullet~ if ${\mathcal C}_1,\ldots,{\mathcal C_t}$ are the open connected components
of \[{\mathbb R}^{n+1}\backslash (\LX_{\infty}\cup \LX_{J}),\] then for each $k$ $(1\leq k\leq t)$, for any ${\mathbf u}\in
{\mathcal C}_k$,  
\[\#\pi^{-1}({\bf u})\cap \LX\cap {\mathbb
R}^{n+s+2}\]
is a constant;

%\item
\textbullet~ if ${\mathcal C}_1,\ldots,{\mathcal C_t}$ are the open connected components of
 \[{\mathbb R}^{n+1}\backslash (\LX_{\infty}\cup \LX_{J}\cup {\LX}_{p}),\] 
then for each $k$ $(1\leq k\leq t)$, for any ${\mathbf u}\in {\mathcal C}_k$,  
\[\#\pi^{-1}({\mathbf u})\cap \LX\cap ({\mathbb R}_{>0}^{n+1}\times {\mathbb R}^{s+1})\]
is a constant. 
%\end{itemize}
\end{theorem}

%\begin{notation}
%{\bf Notation 10.}
%\end{notation}
Before we give the definition of data-discriminant, we study the structures of $\LX_\infty$, $\LX_J$, and $\LX_p$ below.

\textbullet~Proposition \ref{sp} shows that the  structure of the likelihood equations forces $\LX_p$ to be contained in the  union of coordinate hyperplanes defined by $\prod_{k=0}^n u_k$.

\textbullet~Proposition \ref{sj} shows that the structure of the likelihood equations forces $\LX_J\backslash \{{\bf 0}\}$ to be a projective variety. 

\textbullet~Similarly as the proof of Proposition \ref{sj}, we can also show that the structure of the likelihood equations forces $\LX_\infty\backslash \{{\bf 0}\}$ to be a projective variety.

\begin{proposition}\label{sp}
%{\bf Proposition 1.}
For any algebraic statistical model $X$, 
\[{\LX}_{p}\subset {\mathcal V}_a(\Pi_{k=0}^nu_k).\]
\end{proposition}

{\bf Proof.}
By  Definition \ref{LE},
 for any $k$ $(0\leq k\leq n)$, 
\[u_k = p_k (\lambda_1+\frac{\partial g_1}{\partial p_1}\lambda_2+\cdots+\frac{\partial g_s}{\partial p_1}\lambda_{s+1}) - F_k.\]
Hence,  
\[u_k \in \langle F_k, p_k\rangle\cap {\mathbb Q}[u_k]\subset \langle F_0, \ldots, F_{n+s+1}, p_k\rangle\cap {\mathbb C}[{\mathbf u}]\]
So 
\[{\mathcal V}_a(\langle F_0, \ldots, F_{n+s+1}, p_k\rangle\cap {\mathbb C}[{\mathbf u}])\subset {\mathcal V}_a(u_k)\]
By the Closure Theorem \cite{CLO2007}, 
\[{\mathcal V}_a(\langle F_0, \ldots, F_{n+s+1}, p_k\rangle\cap {\mathbb C}[{\mathbf u}])=\overline{\pi({\LX}\cap {\mathcal V}_a(p_k))}\]
Therefore, 
\begin{align*}
{\LX}_{p}&=
\overline{\pi({\LX}\cap {\mathcal V}_a(\Pi_{k=0}^np_k))}\\
&=\overline{\pi({\LX}\cap \cup_{k=0}^n{\mathcal V}_a(p_k))}\\
&=\cup_{k=0}^n\overline{\pi({\LX}\cap {\mathcal V}_a(p_k))}\\
&\subset \cup_{k=0}^n{\mathcal V}_a(u_k)\\
&={\mathcal V}_a(\Pi_{k=0}^nu_k). \Box
\end{align*}

\begin{remark}
%{\bf Remark 2.}
Generally, 
${\LX}_{p}\neq {\mathcal V}_a(\Pi_{k=0}^nu_k)$. For example, suppose the algebraic statistical model is ${\mathcal V}_a(p_0-p_1)\cap \Delta_1$.
Then  ${\LX}_{p}=\emptyset\neq {\mathcal V}_a(u_0u_1)$.
\end{remark}

\begin{notation}
%{\bf Notation 11.}
$\DD_p$ denotes the product $\Pi_{k=0}^nu_k$.
\end{notation}

\begin{proposition}\label{sj}
%{\bf Proposition 2.}
For an algebraic statistical model $X$, we have
${\LX}_J\backslash \{{\bf 0}\}$ is a projective variety in ${\mathbb P}^n$, where ${\bf 0}$ is the zero vector $(0, \ldots, 0)$ in ${\mathbb C}^{n+1}$.
\end{proposition}
%\begin{proof}

{\bf Proof.}
By the formulation of the Lagrange likelihood equations, %\ref{hh}
we can prove that ${\mathcal I}(\pi(\LX\cap {\mathcal V}_a(J))$ is a homogeneous ideal by the two basic facts below, which can be proved by  Definition \ref{LE} and basic algebraic geometry arguments.  
%\begin{itemize}
%\item[C1.]

{\bf C1.} For every ${\mathbf u}$ in  $\pi(\LX\cap {\mathcal V}_a(J))$,  each  scalar multiple  %and for any scalar $\alpha\in {\mathbb C^{*}}$,  
$\alpha{\mathbf u}$ is also in  $\pi(\LX\cap {\mathcal V}_a(J))$. %In fact, it is directly checked by the definitions of $F_0$, $\ldots$, $F_{n+s+1}$ and $J$.
%\item[C2.]

{\bf C2.} For any $S\subset {\mathbb C}^{n+1}$, if for any ${\mathbf u}\in S$ and for any scalar $\alpha \in {\mathbb C}$,  $\alpha{\mathbf u}\in S$, then ${\mathcal I}(S)$ is a homogeneous ideal in~${\mathbb Q}[{\mathbf u}]$. %In fact, we only need to prove that if $B\in {\mathcal I}(S)$, then all the homogenous components of $B$ are also living in ${\mathcal I}(S)$. Suppose the total degree of $B$ is $d$. If $d=0$, it is obviously true. Assume it is always true for $d\leq n$. Suppose $d=n+1$. We rewrite $B$ as $B_0+\ldots+B_n +B_{n+1}$, where $B_i$ is the homogeneous component of $B$ with total degree $i$.  For any ${\mathbf u}\in S$ and for any $a\in {\mathbb C}$, we have 
%\begin{align}
%B_0({\mathbf u}) + \ldots+B_{n+1}({\mathbf u})=0
%\end{align}
%\begin{align}
%B_0(a\cdot u_0, \ldots, a\cdot u_n) +\ldots+ B_{n+1}(a\cdot u_0, \ldots, a\cdot u_n)=\\
%B_0({\mathbf u}) + \ldots+a^{n+1}B_{n+1}({\mathbf u})=0
%\end{align}

%Consider $a^{n+1}\cdot$(1) $-$ (2), we get
%\begin{align}
%\Sigma_{k=0}^n(a^{n+1}-a^k)B_k({\mathbf u})=0
%(a^{n+1}-1)B_0(u_0, \ldots, u_n) + \ldots + (a^{n+1}-a^n)B_{n}( u_0, \ldots,  u_n)=0
%\end{align}
%Note that the equation (3) holds for any $a\in {\mathbb C}$. Choose $b\in {\mathbb Q}$ such that 
%\[b^{n+1}-1\neq 0, \ldots, b^{n+1}-b^n\neq 0.\]
%Let $G=\Sigma_{k=0}^n(b^{n+1}-b^k)B_k({\mathbf u})=0$.
%Now we have 
%$G\in {\mathcal I}(S)$ and the total degree of $G$ is not greater than $n$. Hence $(b^{n+1}-1)B_0, \ldots, (b^{n+1}-b^n)B_{n}\in {\mathcal I}(S)$ and then 
%$B_0, \ldots, B_n\in {\mathcal I}(S)$.  Therefore, we also have $B_{n+1}\in {\mathcal I}(S)$ since $B\in {\mathcal I}(S)$.

%\end{itemize}
That means the ideal ${\mathcal I}(\pi(\LX\cap {\mathcal V}_a(J))$ is generated by finitely many homogeneous polynomials $D_1$, $\ldots$, $D_m$. Therefore, $\LX_J={\mathcal V}_a({\mathcal I}(\pi(\LX\cap {\mathcal V}_a(J)))={\mathcal V}_a(D_1, \ldots, D_m)$. So $\LX_J\backslash \{{\bf 0}\}={\mathcal V}(D_1, \ldots, D_m)\subset {\mathbb P}^n$. $\Box$
%\end{proof}

\begin{notation}\label{ddvj}
%{\bf Notation 14.}
For an algebraic statistical model $X$, 
we define the notation $\DD_J$ according to the codimension of ${\LX}_J\backslash \{\bf{0}\}$ in ${\mathbb P}^n$.
%\begin{itemize}
%\item 

\textbullet~If the codimension is $1$, then assume
${\mathcal V}(D_1), \ldots,  {\mathcal V}(D_K)$ are the codimension $1$ irreducible components in the minimal irreducible decomposition of ${\LX}_J\backslash \{\bf{0}\}$ in ${\mathbb P}^n$ and $\langle D_1 \rangle$, $\ldots$, $\langle D_K \rangle$ are radical. 
$\DD_J$ denotes the homogeneous polynomial $\Pi_{j=1}^KD_j$.
%\item 

\textbullet~If the codimension is greater than $1$, then our  convention is to take  $\DD_J =1$. 
%\end{itemize}
\end{notation}

%As what we have said before,  $\LX_{\infty}\backslash \{\bf{0}\}$ is also a projective variety in ${\mathbb P}^n$.
Similarly, we  use the notation $\DD_{\infty}$ to denote the projective variety $\LX_{J}\backslash \{\bf{0}\}$.
 Now we define the ``data-discriminant'' of Lagrange likelihood equations. 

\begin{defn}\label{dd}{\bf (Data-Discriminant)}
%[Data-Discriminant]\label{dd}
%{\bf Definition 4. (Data-Discriminant)}
For a given algebraic statistics model $X$, the homogeneous polynomial $\DD_{\infty}\cdot\DD_{J}\cdot\DD_p$ is said to be the {\em data-discriminant} (DD) of Lagrange likelihood equations of $X$ and denoted by $\DD$.
%The $\textit{u-discriminant}$ or $\textit{data discriminant }$ of
%an algebraic statistical model is defined to be the $u$ where $\#Crit_{X}\left(u\right)\neq
%MLdegree\left(X\right)$ 
\end{defn}

%\begin{proposition}
%{\bf Proposition 1.}
%For a given algebraic statistics model $X$, $\langle \DD \rangle$ is homogenous and radical in ${\mathbb Q}[u_0, \ldots, u_n]$.  
%\end{proposition}

\begin{remark}
%{\bf Remark 3.}
Note that DD can be viewed as a generalization of the ``discriminant'' for univariate  polynomials. So it is interesting to compare DD with border polynomial (BP) \cite{BP2001} and discriminant variety (DV) \cite{DV2005}.    DV and BP are defined for general parametric polynomial systems. 
DD is defined  for  the likelihood equations but can be  generalized to any square and generic zero-dimensional  system. 
%{\color{magenta} Should we also mention GKZ?}
Generally, for any square and generic zero-dimensional  system, ${\mathcal V}_a($DD$) \subset$ DV $\subset$ ${\mathcal V}_a($BP$)$.   Note that due to the special structure of likelihood equations, DD is a homogenous polynomial despite being an affine system of equations. However, generally, DV is not a projective variety and BP is not homogenous. 
\end{remark}
%\begin{example}[Linear Model]\label{linearmodel}
{\bf Example 1 (Linear Model).}
The algebraic statistic model for the four sided die story in  Section \ref{introduction} is given by 
\[X={\mathcal V}(p_0+2p_1+3p_2-4p_3)\cap \Delta_{3}.\] 
The Langrange likelihood equations are the $F_0=0, \ldots, F_5=0$ shown in  Section \ref{introduction}. 
The Langrange likelihood correspondence is $\LX={\mathcal V}_a(F_0, \ldots, F_5)\subset {\mathbb C}^{10}$. If we choose generic  $(u_0, u_1, u_2, u_3)\in {\mathbb C}^4$, $\pi^{-1}(u_0, u_1, u_2, u_3)\cap \LX=3$, namely the ML-degree is $3$. The data-discriminant is the product of $\DD_{\infty}$, $\DD_{p}$ and $\DD_{J}$, where

$\DD_{\infty}=u_0+u_1+u_2+u_3$,
$\DD_{p}=u_0u_1u_2u_3$, and 

 $\DD_{J}=$
  {\scriptsize$441u_0^4+4998u_0^3u_1+20041u_0^2u_1^2+33320u_0u_1^3+19600u_1^4-756u_0^3u_2+
20034u_0^2u_1u_2+83370u_0u_1^2u_2+79800u_1^3u_2-5346u_0^2u_2^2+55890u_0u_1u_2^2+119025u_1^2u_2^2+4860u_0u_2^3+76950u_1u_2^3+18225u_2^4-1596u_0^3u_3-11116u_0^2u_1u_3-17808u_0u_1^2u_3+4480u_1^3u_3+7452u_0^2u_2u_3-7752u_0u_1u_2u_3+49680u_1^2u_2u_3-17172u_0u_2^2u_3+71460u_1u_2^2u_3+27540u_2^3$\\$u_3+2116u_0^2u_3^2+6624u_0u_1u_3^2-4224u_1^2u_3^2-9528u_0u_2u_3^2+15264u_1u_2u_3^2+14724u_2^2u_3^2-1216u_0u_3^3-512u_1u_3^3+3264u_2u_3^3+256u_3^4$}.

By applying the well known partial cylindrical algebraic decomposition (PCAD) \cite{CH1998} method to the data-discriminant above, we get  that for any $(u_0, u_1, u_2, u_3)\in {\mathbb R}_{>0}^4$, 

\textbullet ~if $\DD_{J}(u_0, u_1, u_2, u_3)>0$, then the system of likelihood equations has $3$ distinct real solutions and $1$ of them is positive; 

\textbullet ~if $\DD_{J}(u_0, u_1, u_2, u_3)<0$, then the system of likelihood equations has exactly $1$ real solution and it is positive. %{\color{magenta} This is confusing because it seems like the real root classification is equivalent to the positive root classification.}

The answer above can be verified by the {\tt RealRootClassification} \cite{BP2001, CDMMX2010} command in {\tt Maple 17}.  In this example, the $\DD_{\infty}$ does not effect the number of real/positive solutions since it is always positive when each $u_i$ is positive. However, generally, $\DD_{\infty}$  plays an important role in real root classification. Also remark that the real root classification is equivalent to the positive root classification for this example but it is not true generally (see Example 6). 

%\end{example}
\section{Algorithm}\label{algorithm}
In this section, we discuss how to compute $\DD$. 
We assume that $X$ is a given statistical model,  $F_0, \ldots, F_{n+s+1}$ are defined as in Definition \ref{LE},
 and $J$ is defined as in  Definition~\ref{nddv}. We rename $F_0, \ldots, F_{n+s+1}$ as $F_0, \ldots, F_m$.
Subsection \ref{standardAlgSection} presents the standard elimination algorithm for reference and Subsection \ref{mainResults} presents our main algorithm (Algorithm \ref{interpolation}).
%We assume the readers are familiar with the Gr\"obner Base \cite{CLO2007}.  

%\begin{notation}
%{\bf Notation 15.}
%We rename $F_0, \ldots, F_{n+s+1}$ as $F_0, \ldots, F_m$.% We also rename the   unknowns  $p_0, \ldots, p_n, \lambda_1, \ldots, \lambda_{s+1}$ as $y_0, \ldots, y_n$, $y_{n+1},\ldots, y_m$, denoted by ${\mathbf y}$. 
%\end{notation}

\subsection{Standard Elimination Algorithm}\label{standardAlgSection}
Considering the  data-discriminant as a projection drives a natural algorithm to compute it. This is the  standard elimination algorithm in symbolic computation:

%\begin{itemize}
%\item 
\textbullet~we compute the $\LX_J$ by {\em elimination} and then get $\DD_J$ by the {\em radical equidimensional decomposition} (see Definition 3 in \cite{DV2005}). The algorithm is formally described in the
Algorithm \ref{dxj};
\begin{algorithm}\label{dxj}
\scriptsize
\DontPrintSemicolon
\LinesNumbered
\SetKwInOut{Input}{input}
\SetKwInOut{Output}{output}
\Input{$F_0, \ldots,F_{m}, J$}
\Output{$\DD_J$}
${\mathcal G}_{\bf u}\leftarrow$ the generator polynomial set of the elimination ideal $\langle F_0, \ldots F_{m}, J\rangle\cap {\mathbb Q}[{\mathbf u}]$\;
$\DD_J\leftarrow$  the codimension $1$ component of the equidimensional radical decomposition of $\langle {\mathcal G}_{\bf u}\rangle$\;
{\bf return} $\DD_J$

\caption{DX-J}
\end{algorithm}
%\item 

\textbullet~we compute  $\LX_\infty$ by the Algorithm {\tt PROPERNESSDEFECTS} presented in \cite{DV2005} and  then get  $\DD_\infty$ by the  radical equidimensional  decomposition.
%(see Definition 3 in \cite{DV2005}). 
We omit the formal description of the algorithm.

The previous algorithms in this subsection can not be used to  compute DDs of algebraic statistical models
in a reasonable time, see  Tables 1--2 in  Section \ref{experiment}.
This motivates the  exploration of a more practical method found in the next subsection.
 
\subsection{Probabilistic Algorithm}\label{mainResults}
First, we prepare the lemmas, then we present the main algorithm (Algorithm \ref{interpolation}).

\textbullet~Lemma 1 is used to linearly transform   parameter space. 

\textbullet~Corollary 1 and Lemma 2 are used to compute the totally degree of $\DD_J$. 

\textbullet~Corollary 2 is used  in the sampling for interpolation. %Corollary 2 is used for computing the sample of interpolation. 

%\begin{remark}
%In the discussion below, by saying 
%\begin{center}
%``for {\em random} $(a_1, \ldots ,a_n)$\\$\in {\mathbb C}^{n}$, a statement is true {\em with algebraic probability $1$}'',  
%\end{center}
%we mean 
%\begin{center}
%``for {\em generic} $(a_1, \ldots, a_n)\in {\mathbb C}^{n}$, the statement is true'',
%\end{center}
%namely
%\begin{center}
%``there exists an affine variety $V$ in ${\mathbb C}^{n}$ such that for any $(a_1, \ldots, a_n)\in {\mathbb C}^{n}\backslash V$, the statement is true'' \cite{SW2005}. 
%\end{center}
%\end{remark}
\begin{lemma}\label{coordinate}
%{\bf Lemma 1.}
For any $G\in {\mathbb Q}[{\mathbf u}]$, there exists an affine variety $V$ in ${\mathbb C}^{n}$ such that for any $(a_1, \ldots, a_n)\in {\mathbb C}^{n}\backslash V$,  the total degree of $G$ equals the degree of 
$B(t_0, t_1, \ldots, t_n)$  {\it w.r.t.} to  $t_0$, where 
\[B(t_0, t_1, \ldots, t_n) = G(t_0, a_1 t_0 + t_1, \ldots, a_n t_0 + t_n)\]
\end{lemma}

%\begin{proof}
{\bf Proof.}
Suppose the total degree of $G$ is $d$ and $G_d$ is the homogeneous component of $G$ with total degree $d$.  For any  $(1, a_1, \ldots, a_n)\in {\mathbb C}^{n+1}\backslash {\mathcal V}_a(G_d)$, 
let $B(t_0, t_1, \ldots, t_n)=G(t_0, a_1 t_0 + t_1, \ldots, a_n t_0 + t_n)$. It is easily seen that 
the degree of 
$B$ {\it w.r.t.} $t_0$ equals $d$. $\Box$
%\end{proof}

\begin{corollary}\label{degree1}
%{\bf Corollary 1.}
For any $G\in {\mathbb Q}[{\mathbf u}]$, there exists an affine variety $V$ in ${\mathbb C}^{2n+2}$ such that for any \[(a_0, b_0, \ldots, a_n, b_n)\in {\mathbb C}^{2n+2}\backslash V,\]  the total degree of $G$ equals the degree of $B(t)$
where
\[B(t) = G(a_0 t+b_0, \ldots, a_n t +b_n).\]
\end{corollary}
%\begin{proof}
%Suppose the total degree of $G$ is $d$ and $G_d$ is the homogeneous component of $G$ with total degree $d$.  For any  $(a_0, \ldots, a_n)\in {\mathbb C}^{n+1}\backslash {\mathcal V}(G_d)$ and for any $(b_0, \ldots, b_n)\in {\mathbb C}^{n+1}$, let $B(t)=G(a_0\cdot t+b_0, \ldots, a_n\cdot t +b_n)$. It is easily seen that 
%the degree of 
%$B(t)$ equals $d$.
%\end{proof}

%\begin{corollary}\label{irreducible}
%{\bf Corollary 2.}
%Suppose $G\in {\mathbb Q}[{\mathbf u}]$. If $G$ is a reducible polynomial in ${\mathbb Q}[{\mathbf u}]$, then for random 
 %\[(a_0, b_0, \ldots, a_n, b_n)\in {\mathbb C}^{2n+2},\]  $G(a_0 t+b_0, \ldots, a_n t +b_n)$ is a reducible polynomial in ${\mathbb Q}[t]$ with algebraic probability $1$. 
% \end{corollary}

%\begin{proof}
%Assume that $G=G_1\cdots G_K$ where $G_1, \ldots, G_K$ are irreducible non-constant polynomials in ${\mathbb Q}[u_0, \ldots, u_n]$. 
%Obviously, for any $(a_0, b_0, \ldots, a_n, b_n)\in {\mathbb C}^{2n+2}$, $B(t)=B_1(t)\cdots B_K(t)$
%where
%\[B(t)=G(a_0\cdot t+b_0, \ldots, a_n\cdot t +b_n), B_i(t)=G_i(a_0\cdot t+b_0, \ldots, a_n\cdot t +b_n)\] 
%Suppose $G_d$ is defined as that in the proof of the Lemma \ref{degree1}. 
%By the proof of the Lemma \ref{degree1}, If  $(a_0, \ldots, a_n)\in {\mathbb C}^{n+1}\backslash {\mathcal V}(G_d)$,
 %then $B_1(t), \ldots, B_K(t)$ are non-constant polynomials in ${\mathbb Q}[t]$. That means $B(t)$ is reducible in ${\mathbb Q}[t]$.
%\end{proof}

\begin{lemma}\label{degree2}
%{\bf Lemma 2.}
There exists an affine variety $V$ in ${\mathbb C}^{2n+2}$ such that for any $(a_0, b_0, \ldots, a_n, b_n)\in {\mathbb C}^{2n+2}\backslash V$,  if 
\[\langle A(t) \rangle=\langle F_0(t), \ldots, F_n(t), F_{n+1}, \ldots, F_m, J \rangle\cap {\mathbb Q}[t]\]
where  $F_i(t)$ is the polynomial by replacing $u_i$ with $a_i t+ b_i$ in $F_i$ $(i=0, \ldots, n)$
and \[B(t)=\DD_J(a_0 t+b_0, \ldots, a_n t +b_n),\]
then $\langle B(t)\rangle =\sqrt{\langle A(t)\rangle}$.
\end{lemma}

%\begin{proof}
{\bf Proof.}
By the definition of $\DD_J$ (Notation \ref{ddvj}),   there exists an affine variety $V_1$ such that for any $(a_0, b_0, \ldots, a_n, b_n)\in {\mathbb C}^{2n+2}\backslash V_1$, $\langle B(t)\rangle$  is radical. Thus, we only need to show that there exists an affine variety $V_2$ in ${\mathbb C}^{2n+2}$ such that for any $(a_0, b_0, \ldots, a_n, b_n)\in {\mathbb C}^{2n+2}\backslash V_2$,  ${\mathcal V}_a(\langle B(t)\rangle) ={\mathcal V}_a(\langle A(t)\rangle)$. 

Suppose $\pi_t$ is the canonical projection: ${\mathbb C}\times {\mathbb C}^{m+1}\rightarrow {\mathbb C}$. 
For any \[t^*\in \pi_t({\mathcal V}_a(F_0(t), \ldots, F_n(t), F_{n+1}, \ldots, F_m, J)),\] 
 let $u_i^* = a_i t^*+ b_i $ (for $i=0, \ldots, n$),  then $(u_0^*, \ldots, u_n^*)\in \pi(\LX\cap {\mathcal V}_a(J))$. Hence $\DD_J(u_0^*, \ldots, u_n^*)=0$ and so $B(t^*)=0$. Thus \[B(t)\in {\mathcal I}(\pi_t({\mathcal V}_a(F_0(t), \ldots, F_n(t), F_{n+1}, \ldots, F_m, J)).\] Therefore,
\begin{align*}
{\mathcal V}_a(A(t))&={\mathcal V}_a({\mathcal I}(\pi_t({\mathcal V}_a(F_0(t), \ldots, F_n(t), F_{n+1}, \ldots, F_m, J)))\\
&\subset {\mathcal V}_a(B(t)).
\end{align*}
For any $t^*\in {\mathcal V}_a(\langle B(t)\rangle)$, let $u_i^* = a_i t^*+ b_i $ for $i=0, \ldots, n$, then 
$(u_0^*, \ldots, u_n^*)\in {\mathcal V}_a(\DD_J)\subset \LX_J$. By the Extension Theorem \cite{CLO2007},  
there exists an affine variety $V_2\subset {\mathbb C}^{2n+2}$ such that if $(a_0, b_0, \ldots, a_n, b_n)\not\in V_2$, then 
$(u_0^*, \ldots, u_n^*)\in \pi(\LX\cap \mathcal{V}_a(J))$, thus \[t^*\in \pi_t({\mathcal V}_a(F_0(t), \ldots, F_n(t), F_{n+1}, \ldots, F_m, J))\subset {\mathcal V}_a(A(t)). \Box\]
%\end{proof}

\begin{corollary}\label{csample}
%{\bf Corollary 3.}
There exists an affine variety $V$ in ${\mathbb C}^{n}$ such that for any $(a_1, \ldots, a_n)\in {\mathbb C}^{n}\backslash V$,  if 
\[\langle A(u_0) \rangle=\langle F_0,F_1^* \ldots, F_n^*, F_{n+1}, \ldots, F_m, J \rangle\cap {\mathbb Q}[u_0]\]
where 
$F_i^*$ is the polynomial by replacing $u_i$ with $a_i$ in $F_i$ ($i=1, \ldots, n$)
and \[B(u_0)=\DD_J(u_0, a_1, \ldots, a_n),\]
then $\langle B(u_0)\rangle =\sqrt{\langle A(u_0)\rangle}$.
\end{corollary}

%%%
%%%
%%%
\begin{algorithm}\label{interpolation}
\scriptsize 
\DontPrintSemicolon
\LinesNumbered
\SetKwInOut{Input}{input}
\SetKwInOut{Output}{output}
\Input{$F_0, \ldots,F_{m}, J$}
\Output{ $\DD_J$}

$a_1, \ldots, a_n\leftarrow $LinearOperator$(F_0, \ldots, F_m, J)$\;
  \For {$i$ {\bf from} $1$ {\bf to} $n$}
        {
        $F'_i\leftarrow$ replace $u_i$ in $F_i$ with $a_i u_0 + u_i$}
     $NewSys\leftarrow F_0, F'_1 \ldots,F'_n, F_{n+1}, \ldots, F_m, J$\; 
    $d, d_0, \ldots, d_n\leftarrow $Degree$(NewSys)$   

 \For {$j$ {\bf from} $1$ {\bf to} $d$}
{

Rename all the monomials of the set
\[\{u_1^{\alpha_1}\cdots u_n^{\alpha_n}|\alpha_1+\ldots+ \alpha_n = j, 0\leq \alpha_i\leq d_i\}\]
as ${U}_{j, 1}, \ldots, {U}_{{j, N_j}}$}

$N\leftarrow \max(N_1, \ldots, N_d)$\;
\For {$k$ {\bf from} $1$ {\bf to} $N$}
{
$b_{ k, 1}, \ldots, b_{k, n}\leftarrow$ random integers\;
$A(u_0)\leftarrow$Intersect$(NewSys, b_{k, 1}, \ldots, b_{k, n})$\;
$C^*_{d, k}, \ldots, C^*_{1, k}\leftarrow$ the coefficients of $A(u_0)$ {\it w.r.t} $u_0^0, \ldots, u_0^{d-1}$\;
}
\For {$j$ {\bf from} $1$ to $d$}
{
${\mathcal M}_j\leftarrow $ $N_j\times N_j$ matrix whose $(k, r)$-entry  is  $U_{j, r}(b_{ k, 1}, \ldots, b_{k, n})$\;
$C_j\leftarrow ({U}_{j, 1}, \ldots, {U}_{{j, N_j}}){\mathcal M}_j^{-1}(C^*_{j, 1}, \ldots, C^*_{j, N_j})^{T}$

}
$\DD_J\leftarrow$ replace $u_1, \ldots, u_n$ in $u_0^d + \Sigma_{i=0}^{d-1}C_{d-i}u_0^i$ with $u_1-a_1\cdot u_0, \ldots, u_n - a_n\cdot u_0$\;
{\bf Return}  $\DD_J$
\caption{({\bf Main Algorithm}) InterpolationDX-J}
\end{algorithm}
 We show an example to explain the basic idea of the probabilistic algorithm and how the lemmas work in the algorithm. 
 
\noindent
{\bf Example 2 (Toy Example for Interpolation Idea).}
Suppose the radical of the elimination ideal $\langle F, J \rangle\cap {\mathbb Q}[{\bf u}]$ is generated by $D(u_0, u_1, u_2)$,  where  $F=u_0p^2 + u_1p + u_2$ and $J=2u_0p+u_1$. We already know that $D$ is homogenous and equals $u_1^2-4u_0u_2$. 
Rather than by the standard elimination algorithm, we compute $D$ by the steps below. 

\textbullet~First, we substitute $u_0=t+11, u_1=3t+2$ and $u_2=5t+6$ into $F$ and $J$ (the integers $1, 11, 3, 2, 5$ and $6$ are randomly chosen). We  compute the radical of the elimination ideal $\langle F(t, p), J(t, p) \rangle\cap {\mathbb Q}[t]$  and get $\langle11t^2+232t+260\rangle$. By  Lemma \ref{degree2}, $D(t+11, 3t+2, 5t+6)=11t^2+232t+260$. By  Corollary  \ref{degree1}, the total degree of $D$ is $2$  (it geometrically means the random line $u_0=t+11, u_1=3t+2, u_2=5t+6$ intersect our desired hypersurface at $2$ points in the parameter space and it is exactly the definition of the degree of hypersurface).  Similarly, we compute the degree of $D$ {\it w.r.t} $u_0, u_1$ and $u_2$ and get $1, 2$ and $1$, respectively.  So all the possible monomials in $D$ are $u_1^2, u_0u_1, u_1u_2$ and $u_0u_2$. 

\textbullet~Assume $D=u_1^2 + (C_1u_0 + C_2u_2)u_1 + C_3u_0u_2$.  We first substitute $u_0=13$ and $u_2=4$ into $F$ and $J$. 
We  compute the radical of the elimination ideal $\langle F(u_1, p), J(u_1, p) \rangle\cap {\mathbb Q}[u_1]$  and get $\langle u_1^2-208 \rangle$.
By  Corollary \ref{csample}, $D(13, u_1, 4)$ equals $u_1^2-208$. Hence,  $13C_1+4C_2=0$ and  $52C_3=-208$. Therefore, $C_3=-4$. We need one more evaluation to solve $C_1$ and $C_2$. So we substitute $u_0=7$ and $u_2=3$ into $F$ and $J$.  Similarly, we get  $7C_1 + 3C_2=0$ and thus $C_1=C_2=0$. Therefore, $D=u_1^2-4u_0u_2$ (the integers $13, 4, 7, 3$ are randomly chosen).  

This example is ``nice''. Because the degree of $D$ {\it w.r.t} $u_1$ equals the total degree of $D$. In general case, if there is no $u_i$ such that the degree of $D$ {\it w.r.t} $u_i$ equals the total degree, then we should apply the linear transformation to change the parameter coordinates before interpolation. Lemma \ref{coordinate} guarantees the linear transformation makes sense.  
  
%\begin{lemma}\label{recover}
%Suppose $G$ is an irreducible homogenous polynomial with total degree $d$ in ${\mathbb Q}[u_0, u_1]$ and the coefficient of $u_0^d$ is $1$.
%For any $a\in {\mathbb Q}$, 
%if \[G(u_0, a)= \Sigma_{i=0}^dc_iu_0^i,\]
%then 
%\[G(u_0, u_1)=\Sigma_{i=0}^d\frac{c_i}{a^{d-i}}u_0^iu_1^{d-i} \]

%\end{lemma}

\begin{algorithm}\label{sample}
\scriptsize
\DontPrintSemicolon
\LinesNumbered
\SetKwInOut{Input}{input}
\SetKwInOut{Output}{output}
\Input{  $F_0, \ldots,F_{m}, J$ and integers $b_1, \ldots, b_n$}
\Output{ $\DD_J(u_0, b_1, \ldots, b_n)$}
\For {$i$ {\bf from} $1$ {\bf to} $n$} {
$F_i^*\leftarrow$ replace $u_i$ in $F_k$ with $b_i$\; 
}

$A(u_0)\leftarrow$ the generator of the elimination ideal $\langle F_0, F_{1}^*,\ldots,F_n^*, F_{n+1}, \ldots F_{m}, J\rangle\cap {\mathbb Q}[u_0]$\;
$A(u_0)\leftarrow$ the monic generator of $\sqrt{\langle A(u_0)\rangle}$\;
{\bf return} $A(u_0)$
\caption{Intersect}
\end{algorithm}

\begin{algorithm}\label{linear}
\scriptsize
\DontPrintSemicolon
\LinesNumbered
\SetKwInOut{Input}{input}
\SetKwInOut{Output}{output}
\Input{  $F_0, \ldots,F_{m}, J$}
\Output{ $a_1, \ldots, a_n$ such that  the total degree of $\DD_J$ equals the degree of $\DD_J(u_0, a_1\cdot u_0 + u_1, \ldots, a_n\cdot u_0 + u_n)$ {\it w.r.t} $u_0$}
$d, d_0, \ldots, d_n\leftarrow $Degree$(F_0, \ldots, F_m, J)$\;
\eIf{$d=d_0$}
{{\bf return } $0, \ldots, 0$}
{\Repeat{$d=d_0$}
   {
    \For {$i$ {\bf from} $1$ {\bf to} $n$}
        { $a_i\leftarrow$ a random integer\; 
        $F'_i\leftarrow$ replace $u_i$ in $F_i$ with $a_i\cdot u_0 + u_i$}
    $NewSys\leftarrow F_0, F'_1 \ldots,F'_n, F_{n+1}, \ldots, F_m, J$\;    
    $d, d_0, \ldots, d_n\leftarrow $Degree$(NewSys)$   
     }
  }
  {\bf return} $a_1, \ldots, a_n$
\caption{LinearOperator}
\end{algorithm}
Now we are prepared to introduce the probabilistic algorithm for computing the $\DD_J$.  We explain the main algorithm (Algorithm \ref{interpolation}) and all the sub-algorithms (Algorithms 4--6) below.

{\bf Algorithm \ref{degree} (Degree).}  The probabilistic algorithm terminates correctly by  Corollary \ref{degree1}
 and Lemma~\ref{degree2}.

 {\bf Algorithm \ref{linear} (LinearOperator).}  The probabilistic algorithm terminates correctly by  Lemma \ref{coordinate}.
%By the Lemma \ref{coordinate}, we develop an probabilistic algorithm to find a proper set of integer ${a_1, \ldots, a_n}$ such that 
%the total degree of $\DD_J(u_0, \ldots, u_n)$ equals the degree of $\DD_J(t_0, a_1\cdot t_0 + t_1, \ldots, a_n\cdot t_0 + t_n)$ {\it w.r.t} $t_0$. 
%That means we always get a ``nice'' $\DD_J$ by applying the linear transformation $u_0=t_0, u_1=a_1\cdot t_0+t_1, \ldots, u_n = a_n\cdot t_0 + t_n$ to the given Lagrange likelihood equations. 

{\bf Algorithm \ref{sample} (Intersect).} The probabilistic algorithm terminates correctly  by  Corollary \ref{csample}.
%By the Corollary \ref{sample}, we develop an probabilistic algorithm to compute the evaluated polynomial $\DD_J(u_0, a_1, \ldots, a_n)$ for 
%random integers $a_1, \ldots, a_n$ without knowing the polynomial $\DD_J$. 

{\bf Algorithm \ref{interpolation} (InterpolationDX-J).} 
%\begin{itemize}
%\item 

{\bf Lines 1--5.} We compute the total degree of $\DD_J$ and the degrees of $\DD_J$ {\it w.r.t} $u_0, \ldots, u_d$: $d, d_0, \ldots, d_n$   by  Algorithm \ref{degree}.  Algorithm \ref{linear} guarantees  that $d_0=d$ by applying a proper linear transformation $u_1=a_1\cdot u_0+u_1, \ldots, u_n = a_n\cdot u_0 + u_n$. 

%\item 
{\bf Lines 6--7.}
 Suppose $\DD_J=u_0^{d}+C_1u_0^{d-1}+\ldots+C_{d-1}u_0+C_{d}$ where $C_1, \ldots, C_{d}\in {\mathbb Q}[u_1, \ldots, u_n]$ and the total degree of $C_j$ is $j$. 
For $j=1, \ldots, n$, we estimate all the possible monomials of $C_j$  by computing the set
\[\{u_1^{\alpha_1}\cdots u_n^{\alpha_n}|\alpha_1+\ldots+ \alpha_n = j, 0\leq \alpha_i\leq d_i\}\] 
Assume the cardinality of the set is $N_j$ and rename these monomials as ${U}_{j, 1}, \ldots, {U}_{{j, N_j}}$. Then we assume 
\[C_j = c_{j, 1}U_{j, 1}+\ldots+c_{j,N_j}U_{j, N_j}\]
where $c_{j, 1}, \ldots, c_{j, N_j}\in {\mathbb Q}$. The rest of the algorithm is to compute  $c_{j, 1}, \ldots, c_{j, N_j}$. 
%\item 

{\bf Lines 8--12.} For each $j$, for $k=1,\ldots, N_j$, for a random integer vector ${\bf b}_k = (b_{k, 1}, \ldots, b_{k, n})$, 
we compute $\DD_J(u_0, {\bf b}_k)$ by Algorithm \ref{sample}.  That means to compute the function value $C_j({\bf b}_k)$ without knowing $
\DD_J$. 
%\item

 {\bf Lines 13--15.} For each $j$, we solve a square linear equation system for the unknowns $c_{j, 1}, \ldots, c_{j, N_j}$:
\begin{align*}
c_{j, 1}U_{j, 1}({{\bf b}_{k}})+\ldots+c_{j,N_j}U_{j, N_j}({\bf b}_{k})=C_j({\bf b}_k), 
\\(k=1, \ldots, N_j)
\end{align*}
It is known that we can choose nice ${\bf b}_k$ probabilistically such that the coefficient matrix of the linear equation system is non-singular. 
%\item 

{\bf Lines 16.} We apply the inverse linear transformation in the parameter space to get the $\DD_J$ for the original $F_0, \ldots, F_m$.
%\end{itemize}

We can also apply the interpolation idea to  Algorithm {\tt PROPERNESSDEFECTS} \cite{DV2005} and get a probabilistic algorithm to compute the $\DD_\infty$.  We omit the formal description of the algorithm. 
\begin{algorithm}\label{degree}
\scriptsize
\DontPrintSemicolon
\LinesNumbered
\SetKwInOut{Input}{input}
\SetKwInOut{Output}{output}
\Input{  $F_0, \ldots,F_{m}, J$}
\Output{ $d, d_0, \ldots d_n$, where $d$ is the total degree of  $\DD_J$ and $d_i$ is the degree of $\DD_J$ {\it w.r.t} each $u_i$ $(i=0, \ldots, n)$}
\For {$i$ {\bf from} $0$ {\bf to} $n$} {

$F_0^*, \ldots, F_n^*\leftarrow$ replace $u_0, \ldots, u_{i-1}, u_{i+1}, \ldots, u_n$ in $F_0, \ldots, F_n$ with random integers\; 

$A(u_i)\leftarrow$ the generator of the elimination ideal $\langle F_0^*, \ldots,F_n^*, F_{n+1}, \ldots F_{m}, J\rangle\cap {\mathbb Q}[u_i]$\;
$A(u_i)\leftarrow$ the generator of $\sqrt{\langle A(u_i)\rangle}$\;
$d_i\leftarrow$ degree of $A(u_i)$\;
$a_i,  b_i\leftarrow$ random integers\;

}
$F_0(t), \ldots, F_n(t)\leftarrow$ replace $u_0, \ldots, u_n$ with $a_0\cdot t + b_0, \ldots, a_n\cdot t+b_n$ in $F_0, \ldots, F_n$\; 
$A(t)\leftarrow$ the generator of the elimination ideal $\langle F_0(t), \ldots,F_n(t), F_{n+1}, \ldots F_{m}, J\rangle\cap {\mathbb Q}[t]$\;
$A(t)\leftarrow$ the generator of $\sqrt{\langle A(t)\rangle}$\;
$d\leftarrow$ degree of $A(t)$\;
{\bf return} $d, d_0, \ldots, d_n$
\caption{Degree}
\end{algorithm}

\begin{remark}
According to the Notation \ref{ddvj},  when the codimension of ${\LX}_J\backslash \{\bf{0}\}$ (${\LX}_{\infty}\backslash \{\bf{0}\}$) is greater than $1$, we define $
\DD_J$ ($
\DD_{\infty}$) is $1$.
Therefore, it is no more true that the number of real/positive solutions still remains constant over the region determined by the data-discriminant. That means if the output of the Algorithm \ref{interpolation} is $1$, we should  use the standard method (elimination or computing Gr\"obner base). 
\end{remark}
\section{Experimental Timings}\label{experiment}

We have implemented the probabilistic algorithm in {\tt Ma-caulay2}. We have also implemented the standard algorithm in {\tt Macaulay2} to do comparisons (Tables 1 and 2). 
Some of the necessary implementation details are shown~below. 
%\begin{itemize}
%\item 

\textbullet~In the Algorithm \ref{dxj}. Line 1,  Algorithm \ref{sample}. Line 3 and Algorithm \ref{degree}. Lines 3 and 8,  we use the  {\tt Macaulay2} command {\tt eliminate} to compute the elimination ideals.  

%\item 
%\textbullet~In the probabilistic algorithm, if $\DD_J$ (or $\DD_\infty$) is reducible, then by the Corollary 2 %\ref{irreducible}  
%we can compute the total degree for each factor and interpolate factor by factor.  

%\item 
\textbullet~The probabilistic algorithm is implemented in two different ways.  The first implementation is to interpolate at once, which is exactly the same as the Algorithm \ref{interpolation}.  
The second implementation is to interpolate step by step. For example, suppose the $\DD_{J}$ is a polynomial in $u_0, u_1, u_2$ and $u_3$, we first compute $\DD_{J}(u_0, u_1, u_2^*, u_3^*)$ by interpolation for some chosen integers $u_2^*$ and $u_3^*$. And then we compute  $\DD_{J}(u_0, u_1, u_2, u_3^*)$ by interpolation.  At this time, it is easy to recover $\DD_{J}$ since $\DD_{J}$ is homogeneous. The algorithm is naive to describe so we omit the formal description.   
%\end{itemize}

We run  Algorithms \ref{dxj} and \ref{interpolation} for many examples to set benchmarks by a 3.2 GHz Inter Core i5 processor (8GB total memory) under OS X 10.9.3. 
There are two kinds of benchmarks, the random models and literature models. 
%\begin{itemize}
%\item 

\textbullet~%As we know,  the likelihood equations of  statistical models  are usually large. 
We generate $2$ groups of ``random models''.  The first group of random models are generated as follows. We first generate a random homogenous polynomial in $3$ variables $p_0, p_1$ and $p_2$ with total degree $2$.   Suppose this homogenous polynomial is a model variant. %We create the likelihood equations according to the Definition 2.
 We repeat the process for $10$ rounds and get $10$ random models. We call this group of $10$ models $2\deg$-models. Similarly, we generate the group of $3\deg$-models. The Table 1 provides the timings of Algorithm \ref{dxj} and Algorithm \ref{interpolation} (with two different implementations) for $2\deg$-models and $3\deg$-models.  

\textbullet~The literature models are the examples presented in the literatures \cite{SAB2005, DSS2009, EJ2014}. Table 2 provides the timings of Algorithm \ref{dxj} and Algorithm \ref{interpolation} (with two different implementations) for the literature models.  For  Examples 3--5 in the Table 2,  the model invariants for these models are list below. Example 6 is given in  Section \ref{rrc}. 
%\begin{example}[Random Censoring Model]\cite{DSS2009}\label{example2}

\noindent
{\bf Example 3 (Random Censoring  (Example 2.2.2 in \cite{DSS2009})).}
\[2p_0p_1p_2 + p_1^2p_2 + p_1p_2^2 - p_0^2p_{12} + p_1p_2p_{12}\]
%\end{example}
%\begin{example}[$3\times 3$ Zero-Diagonal Matrix \cite{EJ2014}]

\noindent
{\bf Example 4 ($3\times 3$ Zero-Diagonal Matrix \cite{EJ2014}).}
 {\footnotesize \begin{align*}
\det \left[
\begin{array}{ccc}   
    0&    p_{12}    & p_{13} \\   
    p_{21} &    0 & p_{23}\\   
    p_{31} &   p_{32} &  0
\end{array}
\right]
\end{align*}}
%\end{example}
%\begin{example}[Grassmannian of $2$-planes in ${\mathbb C}^4$]\cite{SAB2005, EJ2014}

\noindent
{\bf Example 5 (Grassmannian of $2$-planes in ${\mathbb C}^4$ \cite{SAB2005, EJ2014}).}
\[p_{12}p_{34}-p_{13}p_{24}+p_{14}p_{23}\]
%\end{example}
%\end{itemize}

In the Tables 1--2, the columns ``Algorithm 1'' give the timings of Algorithm \ref{dxj}. The columns ``Algorithm 2'' give the timings of Algorithm \ref{interpolation}, where ``S1" and ``S2" means the first and second implementations, respectively.  The red data means the computation has not finished and received no output.  It is seen from the tables that 
%\begin{itemize}
%\item 

\textbullet~for all the benchmarks we have tried, the  Algorithm \ref{interpolation} is more efficient than  Algorithm \ref{dxj};
%\item 

\textbullet~for the random models and  Example 3,  the two implementations of Algorithm \ref{interpolation} have almost the same~efficiency; 
%\item 

\textbullet~for Examples 4--6, the second implementation (interpolation step by step)  of Algorithm \ref{interpolation} is more efficient than the first implementation (interpolation at once). In fact, it takes the same time for the two implementations to get sample points. But it takes more time for the first implementation to compute the inverse of ${\mathcal M}_j$ in   Algorithm \ref{interpolation}. Line 13, which is a  large size matrix with rational entries. 

\textbullet~for  Example 6, with the standard elimination algorithm, our computer  runs out of memory after $12$ days. 
%\end{itemize}

Note that for each  benchmark,   the output of  Algorithm \ref{interpolation} is the same as Algorithm \ref{dxj} when both algorithms  terminate.

%A\ref{interpolation}
%Algorithm \ref{dxj}
\begin{table}\label{table1}
\tiny
\centering
{\footnotesize\caption{Timings of  Computing $\DD_{J}$ for Random Models (s: seconds; h: hours;  S1: Strategy 1; S2: Strategy 2)}}
\begin{tabular}{|c|c|c|c|c|c|} 
\hline
\multicolumn{3}{|c|}{{\bf 2deg-models}}&\multicolumn{3}{|c|}{{\bf 3deg-models}}\\
\hline
 \multirow{2}{*}{{\bf Algorithm \ref{dxj}}}&
\multicolumn{2}{|c|}{{\bf Algorithm \ref{interpolation}}}& 
\multirow{2}{*}{{\bf Algorithm \ref{dxj}}}&
\multicolumn{2}{|c|}{{\bf Algorithm \ref{interpolation}}}\\
\cline{2-3}
\cline{5-6} 
 &{\bf S1}&{\bf S2}&&{\bf S1}&{\bf S2}\\ \hline
4.9s&0.8s &0.6s&\color{red}>2h&800.4s &901.2s\\ \hline
 3.0s &0.7s &0.6s&\color{red}>2h&777.3s &871.5s\\ \hline
  5.0s &0.8s &0.6s&\color{red}>2h&1428.9s &1499.5s\\ \hline
 5.4s &0.8s &0.7s&\color{red}>2h& 1118.9s&1192.9s\\\hline
 6.3s& 0.8s&0.7s&\color{red}>2h& 448.9s&489.8s\\ \hline
 3.9s & 0.7s&0.6s&\color{red}>2h & 1279.6s&1346.1s\\ \hline
  2.0s & 0.7s&0.5s&\color{red}>2h &1286.5s &1409.0s\\ \hline
  1.7s&0.7s &0.5s&\color{red}>2h &1605.9s &1620.9s\\\hline
 3.8s& 0.8s&0.6s&\color{red}>2h & 1099.4s&1242.6s\\ \hline
 5.8s&0.8s &0.7s&\color{red}>2h & 1229.0s&1288.7s\\
\hline\end{tabular}
\end{table}
%\begin{table}
%\centering
%\caption{Timings for  (3, 3)-Random Models}
%\begin{tabular}{|c|c|c|} \hline
% \multirow{2}{*}{Algo. \ref{dxj}}&
%\multicolumn{2}{|c|}{Algo. \ref{interpolation}}\\
%\cline{2-3} 
% &Strat. 1&Strat. 2\\ \hline
%>2h&800.4s &901.2s\\ \hline
%>2h&777.3s &871.5s\\ \hline
%>2h&1428.9s &1499.5s\\ \hline
%>2h& 1118.9s&1192.9s\\\hline
%>2h& 448.9s&489.8s\\ \hline
%>2h & 1279.6s&1346.1s\\ \hline
%>2h &1286.5s &1409.0s\\ \hline
%>2h &1605.9s &1620.9s\\\hline
%>2h & 1099.4s&1242.6s\\ \hline
%>2h & 1229.0s&1288.7s\\
%\hline\end{tabular}
%\end{table}

\begin{table}\label{table2}
\tiny
\centering
\caption{Timings of Computing $\DD_{J}$ for Literature Models (s: seconds; h: hours; d: days;  S1: Strategy 1; S2: Strategy 2)}
\begin{tabular}{|c|c|c|c|} \hline
\multirow{2}{*}{{\bf Models}}& \multirow{2}{*}{{\bf Algorithm \ref{dxj}}}&
\multicolumn{2}{|c|}{{\bf Algorithm \ref{interpolation}}}\\
\cline{3-4} 
 &&{\bf S1}&{\bf S2}\\ \hline
{\bf Example 3} &11.1s&5.3s &6.4s\\ \hline
{\bf Example 4} &36446.4s &360.2s &56.3s\\\hline
{\bf Example 5} &{\color{red}>16h}  &{\color{red}>16h} &2768.2s\\ \hline
{\bf Example 6} &{\color{red}>12d}&\color{red}>30d &30d\\
\hline\end{tabular}
\end{table}

\section{Conclusions and last example}\label{conclusion}
In order to classify the data according to  the number of real/positive solutions of likelihood equations, we study the data-discriminant and develop a probabilistic algorithm to compute it. 
Experiments show that the probabilistic algorithm is more practical than the standard elimination algorithm. 
This is our first application of real root classification method on the MLE/likelihood equations problem. 
 Our future work aims to 
%\begin{itemize}
%\item 

\textbullet~improve  Algorithm \ref{interpolation} (note that  Algorithm \ref{interpolation} is applying evaluation/interpolation technique to the standard method. It is not the first time that such an approach is investigated. In \cite{GLS2000, Schost2003}, Newton--Hensel lifting has been applied to compute (parametric) geometric resolutions.  It is hopeful that Algorithm \ref{interpolation} will be more powerful if we apply the Newton-Hensel lifting techniques to balance the time consuming of the evaluation and lifting steps);

\textbullet~study the data-discriminants of different formulations of likelihood equations for the same algebraic statistical model %and define ``data-discriminant'' for the algebraic statistical model;
%\item 

%\textbullet~develop special algorithms to compute data-discriminants for some special structured models; 
%\item 

\textbullet~develop algorithms for computing real root classification for likelihood equations. 

More broadly, the ideas in Subsection \ref{mainResults} and  Algorithm \ref{interpolation} can be applied to compute discriminants when the Newton polytope is known. 

\subsection{{\large${3\times 3}$} symmetric matrix model}\label{rrc}
We end the paper with the discussion of real root classification on the $3\times 3$ symmetric matrix model.

Consider  the following story with dice. A gambler has a coin,
 and two pairs of three-sided dice. The coin and the dice are all unfair. However, the two dice in the same pair have the same weight.  He plays the same game $1000$ rounds. In each round, he first tosses the coin.  If the coin lands on side $1$,  he tosses the first pair of dice. If the coin lands on side $2$, he tosses the second pair of dice. After the $1000$ rounds,  he records a $3\times 3$ data matrix $[\overline{u}_{ij}]$ $(i, j=1, 2, 3)$ where $\overline{u}_{ij}$ is the 
 the number of times for him to get the sides $i$ and $j$ with respect to the two dice.  By the matrix $[\overline{u}_{ij}]$, he is trying to estimate the probability $\overline{p}_{ij}$ of getting the sides $i$ and $j$ with respect to the two dice. 
 
% As mentioned in the introduction, one approach the gambler can try is to find the $\hat{p}_{ij}$ that maximizes the so-called {\em likelihood function} $\frac{\Pi \overline{p}_{ij}^{\overline{u}_{ij}} }{(\Sigma \overline{p}_{ij})^{\Sigma \overline{u}_{ij}}}$ subjected to some reasonable constraints. It is obviously seen that one of the constraints should be $\Sigma \overline{p}_{ij}=1$. However, there can be more implicit constraints.  
 %Assume that the probabilities of observing the sides $1$ and $2$ of the coin are $c_1$ and $c_2$, and the probabilities of observing the sides $1, 2$ and $3$ of one die in the first and second pair are $[b_1, b_2, b_3]$ and $[r_1, r_2, r_3]$, respectively. We know
It is easy to check that the matrix 
 {\footnotesize\begin{align*}
\left[
\begin{array}{ccc}   
    \overline{p}_{11} &    \overline{p}_{12}    & \overline{p}_{13} \\   
    \overline{p}_{21} &    \overline{p}_{22}   & \overline{p}_{23}\\   
    \overline{p}_{31} &   \overline{p}_{32} &  \overline{p}_{33} 
\end{array}
\right]
\end{align*}}
%is equal to
%\begin{align*}
%c_1\left[
%\begin{array}{c}   
   % b_1  \\   
    %b_2 \\   
    %b_3 
%\end{array}
%\right][b_1, b_2, b_3]+c_2\left[
%\begin{array}{c}   
  %  r_1  \\   
   % r_2 \\   
   % r_3 
%\end{array}
%\right][r_1, r_2, r_3].
%\end{align*} }
%Therefore, $\overline{p}_{ij}=\overline{p}_{ji}$ and the matrix $[\overline{p}_{ij}]$ has at most rank $2$.  
\noindent
is symmetric and has at most rank $2$.
Let 
{\footnotesize\begin{equation*}
p_{ij}=
\begin{cases}
\overline{p}_{ij} & i=j\\
\frac{1}{2}\overline{p}_{ij} & i<j
\end{cases}, \;\;
u_{ij}=
\begin{cases}
\overline{u}_{ij} & i=j\\
\overline{u}_{ij}+\overline{u}_{ji} & i<j
\end{cases}.
\end{equation*}}
\noindent
We have an algebraic statistical model below. 

%\begin{example}[$3\times 3$ Symmetric Matrix Model]
\noindent
{\bf Example 6 ($3\times 3$ Symmetric Matrix Model).}
%[$3\times 3$ Symmetric Matrix]
The algebraic statistical model for the dice story  is given~by 
\[X={\mathcal V}(g)\cap \Delta_5,\]
where 
{\footnotesize \begin{equation*}
g=\det\left[
\begin{array}{cccc}   
    2p_{11} &    p_{12}    & p_{13} \\   
    p_{12} &    2p_{22}   & p_{23}\\   
    p_{13} & p_{23} & 2p_{33} 
\end{array}
\right],
\end{equation*}}
\[\Delta_{5}=\{(p_{11},\ldots,p_{33})\in {\mathbb R}_{>0}^{6}|p_{11}+ p_{12}+ p_{13}+ p_{22}+ p_{23}+ p_{33}=1\}.\]
%\end{example}

The gambler's problem is equivalent to 
%\begin{center}
{\bf maximizing the likelihood function
$\frac{\Pi p_{ij}^{u_{ij}} }{(\Sigma p_{ij})^{\Sigma u_{ij}}}$ $(i
\leq  j)$ subjected to ${\mathcal V}(g)\cap \Delta_5$.}
%\end{center}
%\begin{example}\label{model}
According to the Definition 2, we present the Langrange likelihood equations below. 
{\footnotesize  \begin{align*}
F_0&=p_{11}\lambda_1+(8p_{22}p_{33}-2p_{23}^2)p_{11}\lambda_2-  u_{11}=0\\
F_1&=p_{12}\lambda_1+(2p_{13}p_{23}-4p_{12}p_{33})p_{12}\lambda_2 -  u_{12}=0\\
F_2&=p_{13}\lambda_1+(2p_{12}p_{23}-4p_{13}p_{22})p_{13}\lambda_2-  u_{13}=0\\
F_3&=p_{22}\lambda_1+(8p_{11}p_{33}-2p_{13}^2)p_{22}\lambda_2-   u_{22}=0\\
F_4&=p_{23}\lambda_1+(2p_{12}p_{13}-4p_{11}p_{23})p_{23}\lambda_2- u_{23}=0\\
F_5&=p_{33}\lambda_1+(8p_{11}p_{22}-2p_{12}^2)p_{33}\lambda_2- u_{33}=0\\
F_6&=g(p_{11}, p_{12}, p_{13}, p_{22}, p_{23}, p_{33})=0\\
F_7&=p_{11} + p_{12} +p_{13}+p_{22}+p_{23}+p_{33}-1=0
\end{align*}}
where $p_{11}, p_{12}, p_{13}, p_{22}, p_{23}, p_{33}, \lambda_1$ and $\lambda_{2}$ are unknowns and 
$u_{11}$, $u_{12}$, $u_{13}$, $u_{22}$, $u_{23}$ and $u_{33}$ are parameters.
%\begin{itemize} 
   %          \item $p_{11}, p_{12}, p_{13}, p_{22}, p_{23}, p_{33}, \lambda_1, \lambda_{2}$ are unknowns,
      %       \item $u_{11},u_{12}, u_{13}, u_{22}, u_{23},u_{33}$ are parameters.
         %    \end{itemize}
 
We have $8$ equations in $8$ unknowns with $6$ parameters and the ML-degree is $6$  \cite{SAB2005}. 
%We are interested in the RRC problem: {\bf for which $u_{ij}$, the polynomial system has $1, 2, \ldots, 6$
%real/positive solutions? }%\begin{example}
%For the statistical model of the  dice story in the introduction (Example 1), 
%\begin{itemize}
%\item 
 By the Algorithm \ref{interpolation}, we have computed $\DD_J$, which has $1307$ terms with total degree $12$. 
%\item  
By a similar computation, we get $\DD_\infty$\footnote{\scriptsize See  $\DD_J$ and $\DD_\infty$
 % for the $3\times 3$ symmetric matrix model 
 on the second author's website:\\
 sites.google.com/site/rootclassificaiton/publications/DD}whose last 
%$\DD_\infty=(u_{11}+u_{22}+u_{33}+u_{12}+u_{13}+u_{23})
%(u_{11}+u_{22}+u_{12})(u_{11}+u_{33}+u_{13})(u_{22}+u_{33}+u_{23})
%(u_{12}+2u_{22}+u_{23})(u_{13}+2u_{33}+u_{23})(u_{13}+2u_{11}+u_{12})( 8u_{11}u_{22}u_{33}-2u_{11}u_{23}^2- 2u_{12}^2u_{33}+ 2u_{12}u_{13}u_{23}- 2u_{13}^2u_{22})$. 
factor  is exactly $g(u_{11}, \ldots, u_{33})$
%{\footnotesize \begin{align*}
%g(u_{11}, \ldots, u_{33})
%=det\left[
%\begin{array}{cccc}   
 %   2u_{11} &    u_{12}    & u_{13} \\   
  %  u_{12} &    2u_{22}   & u_{23}\\   
   % u_{13} & u_{23} & 2u_{33} 
%\end{array}
%\right]
%\end{align*}}
and all the other factors are positive when each $u_i$ is positive.  

For the data-discriminant $\DD$ we have computed above, we have also computed\footnote{\scriptsize The sample points were first successfully computed by one of the 
anonymous referees.} at least one rational point (sample point) from each open connected component of $\DD\neq 0$ using  {\tt RAGlib}\cite{SS2003, HS2012, GS2014}.  
With these sample points we can solve the real root classification problem on the open cells. 
By testing all $236$ sample points, we see that  if $g(u_{11}, \ldots, u_{33})$\\$\neq 0$, then\footnote{\scriptsize This proves the real version of the RRC conjecture in the previous version of this manuscript.} 
%\begin{itemize}

~\textendash~if $\DD_J(u_{11}, \ldots, u_{33})>0$, then the system has $6$ distinct real solutions and there can be $6$ positive solution or $2$ positive solutions;

%~~\textendash~if $\DD_J(u_{11}, \ldots, u_{33})=0$, then the system has $4$ distinct
%positive solutions;

~\textendash~if $\DD_J(u_{11}, \ldots, u_{33})<0$, then the system has $2$ distinct
real (positive) solutions.

With $2$ of these sample points, we see that  the sign of $\DD$ is not enough 
to classify  the positive solutions.  
For example, for the sample point  
{\footnotesize $(u_{11}=1, u_{12}=1, u_{13}=\frac{280264116870825}{295147905179352825856}, u_{22}=1, u_{23}=\frac{34089009205592922038535}{141080698675730650759168},$ $u_{33}=\frac{32898355113670387769001}{141080698675730650759168})$}, 
the system has $6$ distinct positive solutions. 
While for the sample point {\footnotesize$(u_{11} = 1, u_{12} = 1, u_{13} = 199008, u_{22} = 30, u_{23} = 2022, u_{33} =1)$}, the system has also
$6$ real solutions but only $2$ positive solutions\footnote{\scriptsize This disproves the positive version of the RRC conjecture in the previous version of this manuscript.}.

%The sample points also show that the open set $\DD_{\infty}\neq 0$ and $\DD_J>0$ has at least $4$ connected components. 

%By the data-discriminant above, we have tested a great amount of data and 
%conclude the conjecture below. It is an experimental answer to the RRC problem.   
%\item  $\LX_{p}={\mathcal V}(u_{11}u_{12}u_{13}u_{22}u_{23}u_{33})$, more specifically, for any $i,j\;(i\leq j)$, 
%\[\overline{\pi({\mathcal LX}\cap {\mathcal V}(p_{ij}))}={\mathcal V}(u_{ij}).\] 
%\end{itemize}
%\end{example}

%\noindent
%{\bf RRC Conjecture for $3\times 3$ Symmetric Matrix Model. }
%For any $(u_{11},\ldots,u_{33})\in {\mathbb R}_{>0}^{6}$, 
%\begin{itemize}
%\item 

%\textbullet~if $g(u_{11}, \ldots, u_{33})=0$, then the system of likelihood equations has exactly $1$ real (positive) solution; 
%\item 

%\textbullet~if $g(u_{11}, \ldots, u_{33})\neq 0$, then 
%\begin{itemize}

%~~\textendash~if $\DD_J(u_{11}, \ldots, u_{33})>0$, then the system has $6$ distinct real (positive) solutions;

%~~\textendash~if $\DD_J(u_{11}, \ldots, u_{33})=0$, then the system has $4$ distinct
%positive solutions;

%~~\textendash~if $\DD_J(u_{11}, \ldots, u_{33})<0$, then the system has $2$ distinct
%real (positive) solutions.
%\end{itemize}
%\end{itemize}

%\end{itemize} 
%\end{document}  % This is where a 'short' article might terminate

%ACKNOWLEDGMENTS are optional
\section{Acknowledgments}
%We would love to thank NIMS for inviting us to join the wonderful Thematic Program 2014 on Applied Algebraic Geometry. 
We  thank  Professors Bernd Sturmfels,   Hoon Hong, Jonathan Hauenstein and  Frank Sottile for their valuable advice. 
We also thank Professors Mohab Safey EI Din and  Jean-Charles Faugere for their  software  advice on {\tt RAGlib} and  {\tt FGb} respectively. 
We also  especially thank the anonymous referees for their   insightful suggestions to greatly improve the~paper.
%Acknowledgments go to Professor Bernd Sturmfels,  Professor Hoon Hong, Professor Jonathan Hauenstein and Professor Frank Sottile for their valuable advice.  

\bibliographystyle{abbrv}

{\small\bibliography{DDManuscript}}

 % sigproc.bib is the name of the Bibliography in this case
% ACM needs 'a single self-contained file'!
%\balancecolumns
\end{document}